\documentclass[preprint,showpacs,aps]{revtex4}
\usepackage{amsmath}
\usepackage{amssymb}
\usepackage{graphicx}
\usepackage{dcolumn}
\usepackage{bm}
\usepackage{float}

\begin{document}

\title{Optical Tamm state and extraordinary light transmission through
nanoaperture}

\author{Ilya V. Treshin}
\author{Vasily V. Klimov}
\email{vklim@sci.lebedev.ru}
\affiliation{P.N. Lebedev Physical Institute, Russian Academy of Sciences,
53 Leninsky Prospekt, Moscow 119991, Russia}

\author{Pavel N. Melentiev}
\author{Victor I. Balykin}
\affiliation{Institute for Spectroscopy Russian Academy of Sciences,
5 Phizicheskaya str., Troitsk, Moscow 142190, Russia}

\begin{abstract}
We investigate the light transmission through a nanoaperture in a metal film
deposited on a planar metamaterial. An effect of an anomalously high light
transmission through the nanoaperture is revealed, which we associate with
the enhancement of the field at the interface of the planar structure
``metamaterial-metal film'' due to the appearance of an optical Tamm state.
In this structure, we also observe an ``optical~diode'' effect:
the light transmission radically changes as the direction of irradiation of
the structure is reversed. Our numerical results agree well
with experimental data.
\end{abstract}

\date{\today}


\maketitle

\section{\label{Section_01} INTRODUCTION}

The effect of an extraordinary transmission of light through an array of
nanoapertures in a metal film has been observed for the first time by
Ebbesen et al.~\cite{ref_01}. The effect describes the light transmission
through a nanoaperture array, which is considerably higher in magnitude than
predicted according to the Bethe-Bouwkamp theory~\cite{ref_02, ref_03}.
At present an extraordinary light transmission has been observed both
for a periodic nanoaperture array in a metal film and for single
nanoapertures in it. In the latter case, various techniques are used
to increase the light transmission. For example, in~\cite{ref_04, ref_05},
a method of structuring of the surface near the aperture is applied in
order to match the wave vectors of the incident electromagnetic field and
plasmons on the film surface. In general, there exists a large variety of
physical mechanisms and methods of their realization that lead to similar
effects. It seems that, in all these cases, the field intensity in the region
of the aperture is enhanced due to the excitation of propagating plasmons,
and precisely this enhancement of the field causes the intensity of
the passing light to increase~\cite{ref_06}. At the same time,
the field enhancement can also occur in ideally conducting geometries, where
there are no plasmons but an anomalously high transmission also
arises~\cite{ref_07, ref_08}.

Recently, an extraordinary increase in the light transmission has been
experimentally detected in the case of a gold film with nanoapertures on
the surface of a multilayer periodic dielectric structure, a planar
metamaterial~\cite{ref_09, ref_10}. At a certain frequency of the incident
light, an \textit{optical Tamm state} of the electromagnetic field
arises~\cite{ref_11,ref_11a}, in which the intensity of the magnetic field
from the side of the metamaterial on the surface of the metal film
has a maximum~\cite{ref_12}.
(In what follows, we will term the light frequency at which the optical Tamm
state arise the resonance frequency.) The light transmission through
nanoapertures increases by several orders of magnitude compared to the case in
which there is no metamaterial.

In view of the importance of the observed effect, the question of its
theoretical explanation arises, to which this work is devoted. The plan of
the remaining part of the paper is as follows. Section~\ref{Section_02}
presents the descriptions of the geometry and model of the numerical
experiment. Results of the numerical simulation of a gold film on the surface
of a metamaterial with and without apertures are presented
in sections~\ref{Section_03} and~\ref{Section_04}, respectively.
Section~\ref{Section_05} is devoted to the discussion of the results.

\section{\label{Section_02} Discussion of the model of the numerical
experiment}

The theoretical calculation of this work is based on the geometry of
experiment of~\cite{ref_09}. The values of the geometric parameters and the
optical properties of materials were taken from that paper, unless otherwise
specified.

\begin{figure}[h!]
\center{\includegraphics[width = 84 mm]{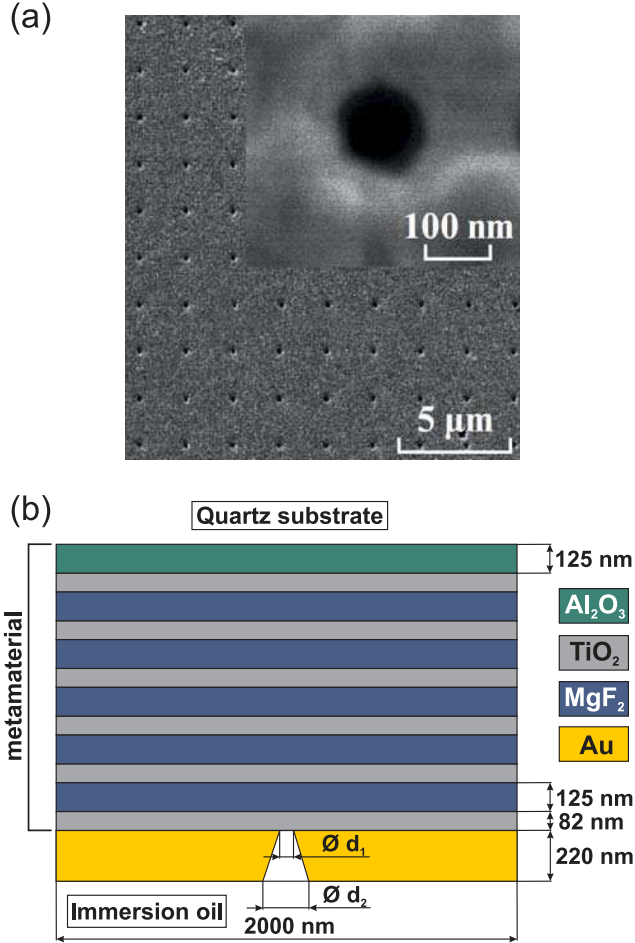}}
\caption{(a) Images of the array and one of the nanoapertures
in the experiment of~\cite{ref_09} obtained by the scanning
electron microscopy method; (b)~schematic image of the cross
section of one period of the nanoaperture array in the gold film
on the metamaterial used in the numerical experiment.}
\label{fig_1}
\end{figure}

The experiment of~\cite{ref_09} is extremely subtle, because it is very
difficult to control the manufacturing process (the shape of apertures, the
thicknesses of layers) with a nanometer accuracy. Figure~\ref{fig_1}(a)~shows
the images of the array and one of nanoapertures used in the experiment
of~\cite{ref_09}.
We can assume from this figure that the nanoaperture does not have an ideal
cylindrical shape. Therefore, in the numerical model, the aperture was
approximated by a generalized truncated cone.

A cross section of one period of the structure ``metamaterial-metal film''
that was used in the numerical experiment is shown in Fig.~\ref{fig_1}(b).

A 12-layer periodic dielectric structure (metamaterial) is deposited on a
quartz substrate (which is shown at the top in~Fig.~\ref{fig_1}(b)).
The dielectric
structure consists of an $\rm Al_{2}O_{3}$ layer (with a thickness of~125~nm)
and alternating $\rm TiO_{2}$ and $\rm MgF_{2}$ layers (with their
thicknesses being~82~and~125~nm, respectively). A gold film with a
thickness of~220~nm is deposited on the surface of the metamaterial (at the
``bottom'' in~Fig.~\ref{fig_1}(b)). Since the thickness of this film
considerably
exceeds the thickness of the skin layer~($\thicksim$~80~nm), the film is
opaque in the considered wavelength range of the incident light
(from 575~to~875~nm). Nanoapertures in the metal film form a square
lattice with a period of~2~$\mu$m. In the numerical model, the aperture is
approximated by a truncated cone, with the diameters of its bases
being $d_{1}$~(on the $\rm TiO_{2}$ layer) and $d_{2}$~(on the opposite
side). The whole space inside the aperture and behind the gold film
is filled with an immersion oil. In~\cite{ref_09}, the thickness of
the $\rm SiO_{2}$ substrate was~2~mm, and the immersion oil was
used to reduce the reflection. In our numerical experiment, the quartz layer
and immersion oil are assumed to be infinitely thick. The refractive
indices~$n$ of all the used dielectric materials were taken from the
experimental data of~\cite{ref_09}: $\rm SiO_{2}$~($n_{q} = 1.443$),
$\rm Al_{2}O_{3}$~($n = 1.63$), $\rm TiO_{2}$~($n = 2.23$),
$\rm MgF_{2}$~($n = 1.38$), and immersion oil~($n_{io} =1.51$).
In calculations, the dispersion of the dielectrics was neglected.
The dispersion dependence of the dielectric permittivity of gold
was taken from~\cite{ref_13}.

The light transmission through nanoapertures was simulated for the normal
incidence of light on the structure. In this case, the wave that was incident
on the structure from the side of the quartz substrate (from above) was
expressed as~$\mathbf{E}_{up}=\mathbf{E}_{0}\exp(ik_{0}n_{q}z)$, while
the wave incident from the side of the immersion oil
was given by~$\mathbf{E}_{down}=\mathbf{E}_{0}\exp(ik_{0}n_{io}z)$.

\section{\label{Section_03} Optical properties of a structure
``metamaterial-metal film''}

Initially, we will consider the optical properties of a
structure~``metamaterial–metal film'' with no apertures. To do this, we will
use the analytical method of~\cite{ref_14, ref_15, ref_16}. In each separate
layer, the solution is represented as a sum of the ``incident'' and
``reflected'' waves. The amplitudes of these waves can be found from the
condition of continuity for the tangential components of the electric and
magnetic fields at the boundary between the layers. As a result, we obtain
a system of linear algebraic equations, the numerical solution of which was
found using~$\rm MATLAB^{\text{\textregistered}}$. Let us consider initially
the case in which the light is incident on the structure~``metamaterial–metal
film'' with no apertures from the side of the quartz [from the top,
see~Fig.~\ref{fig_1}(b)].

\begin{figure}[h!]
\center{\includegraphics[width = 84 mm]{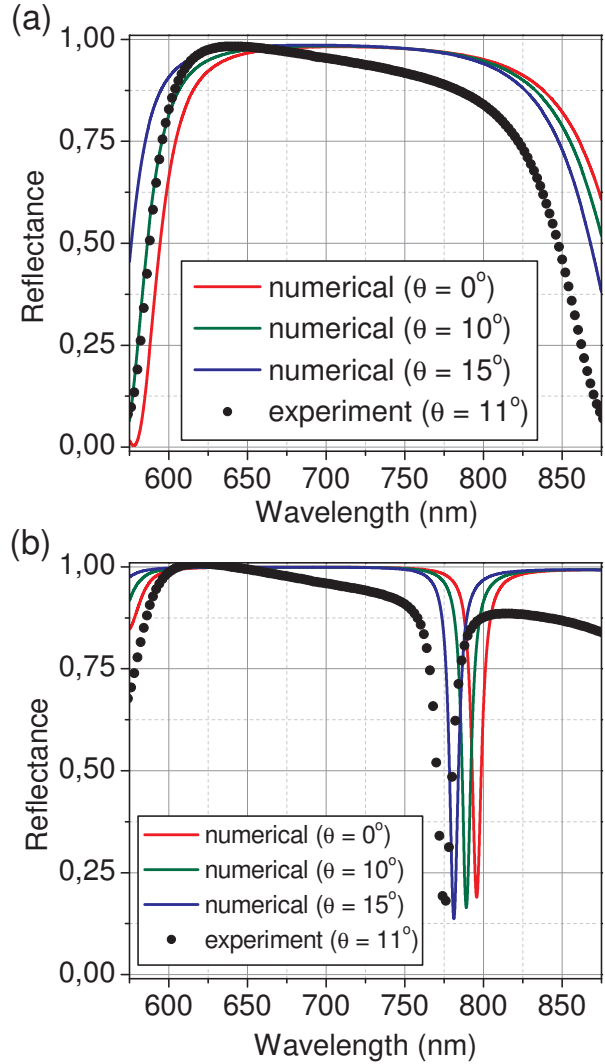}}
\caption{Dependences of the reflection coefficient on the wavelength and on
the angle of incidence of light for (a) the metamaterial and (b) the optical
structure ``metamaterial-metal film''. In the two cases, the light
is incident from the side of the~$\rm SiO_{2}$~layer.}
\label{fig_2}
\end{figure}

Figures~\ref{fig_2}(a)~and~\ref{fig_2}(b)~show the calculated
and experimental~\cite{ref_09}
dependences of the energy reflection coefficient on the wavelength and on the
angle of incidence of light for the metamaterial and the gold film on it,
respectively. The experimental curves are shown in black. It can be seen
from~Fig.~\ref{fig_2}~that there is a qualitative agreement between
the theory and
experiment. In the numerical experiment, we also slightly varied
the thicknesses of the layers, as a result of which the agreement became
much better. This is indicative of possible errors in specifying the optical
constants of the materials or of the imperfect preparation of experimental
specimens.

A characteristic feature of the dependence of the reflection coefficient of
the optical structure ``metamaterial-metal film'' is the occurrence of
a narrow resonance dip. The corresponding resonant wavelength can be
estimated from the condition \cite{ref_12}:

\begin{equation}
\label{(1)} r_{M}r_{MM} = 1,
\end{equation}

where $r_{M}$ is the amplitude reflection coefficient
at~Au-$\rm TiO_{2}$~interface, while $r_{MM}$ is the amplitude reflection
coefficient of the wave incident from $\rm TiO_{2}$ half-space on
the metamaterial starting with a~$\rm TiO_{2}$~layer.

\begin{figure}[h!]
\center{\includegraphics[width = 84 mm]{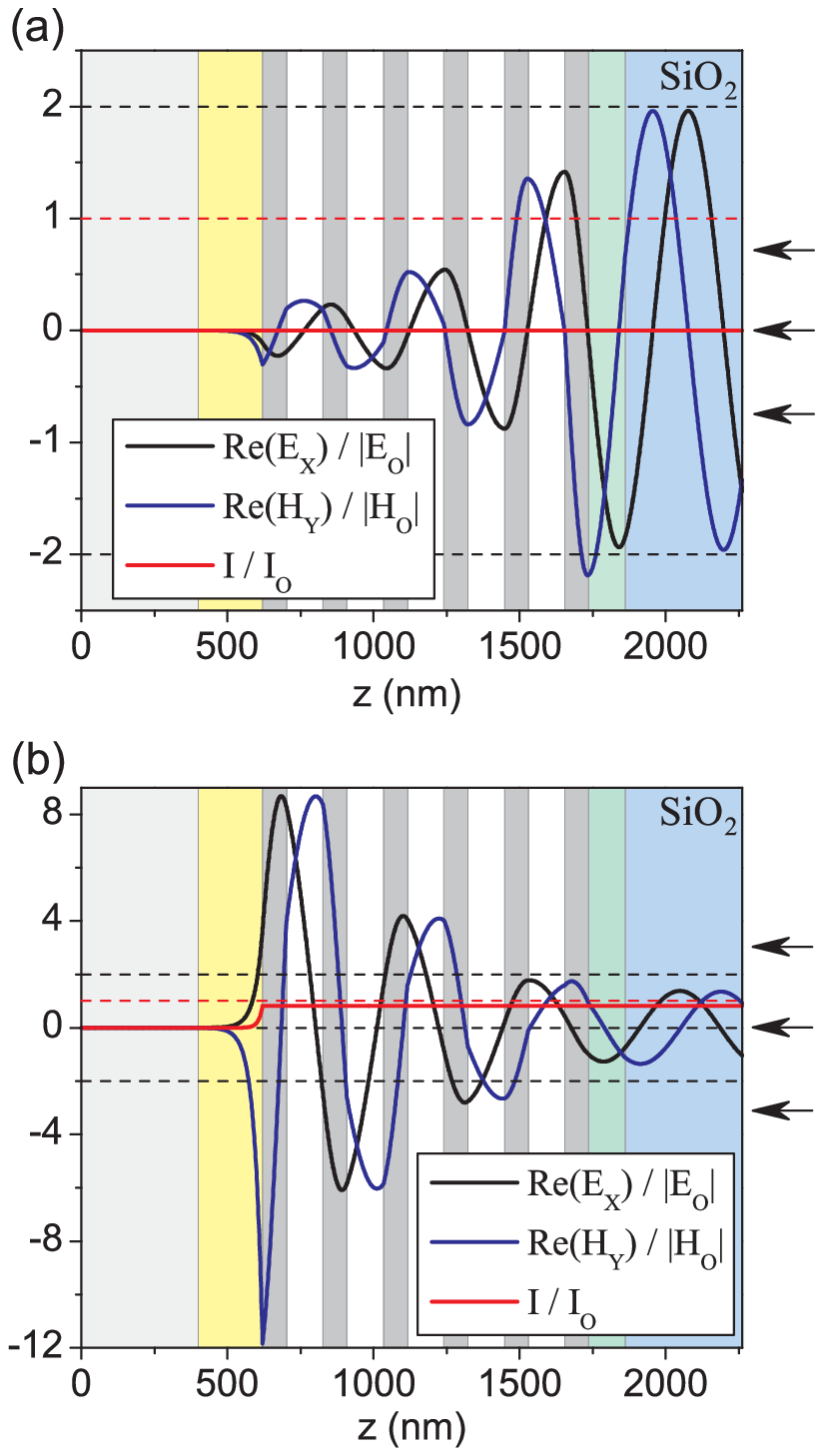}}
\caption{Relative electric (black) and magnetic (blue) field strengths and
Poynting vector flux (red) in the optical
structure~``metamaterial-metal film'' in relation to the $z$ coordinate inside
the structure: (a)~nonresonance case, $\lambda$~=~700~nm; (b)~resonance case,
$\lambda$~=~796~nm. The arrows show the direction of incidence of the light
on the structure. The gold film is shown in yellow, and the metamaterial
is shown by alternating white and gray bars. The dashed lines corresponds
to~1,~$\pm$~2~guidelines.}
\label{fig_3}
\end{figure}

In order to elucidate the physical nature of this narrow resonance peak, we
calculated the distributions of the electric and magnetic fields and of
the energy flux in this structure, Fig.~\ref{fig_3}. The field and
the energy flux
distributions were normalized to the values of ($E_{0}$,~$H_{0}$) in
the incident wave.

From~Fig.~\ref{fig_3}(a), it can be clearly seen that, in the nonresonance
case~($\lambda$~=~700~nm), almost all the energy is reflected on the surface
of the metamaterial, and the amplitudes of the electric and magnetic fields
exponentially decay as the gold film is approached. The energy flux inside
the metamaterial is also close to zero. Conversely, in the resonance case,
i.e., in the case of the appearance of the optical Tamm state,
($\lambda$~=~796~nm,~Fig.~\ref{fig_3}(b)), the reflection is small, while the
intensities of the electric and magnetic fields near the film are high.
The energy flux is directed inward the metamaterial and is absorbed nearly
completely in the gold film. In this case, the transmission coefficient is
very small~($ \thicksim 10^{-6}$). An increase in the amplitudes of
the electric and magnetic fields near the metamaterial surface is caused by
the occurrence of the optical Tamm state of the electromagnetic
field~\cite{ref_11,ref_11a, ref_12}. It is especially important to note that
the magnetic field is maximal at the interface metamaterial–metal, and
it is the increase in this field that is responsible for the considerable
increase in the light transmission through
the nanoaperture~(see section~\ref{Section_05}).

If the structure ``metamaterial-metal film'' is illuminated from the opposite
side, the situation radically changes. In this case, the reflection
coefficient does not have any resonance peak similar to the peak
in~Fig.~\ref{fig_2}(b), since the light does not penetrate into
the metamaterial due
to the absorption in the gold film. Therefore, for the reflection from
the structure ``metamaterial-metal film'', we have an \textit{asymmetric}
situation with respect to the direction of incidence of the light onto the
structure; namely, the \textit{reflection coefficients for the light that is
incident from opposite sides do not coincide with each other}.

\begin{figure}[h!]
\center{\includegraphics[width = 84 mm]{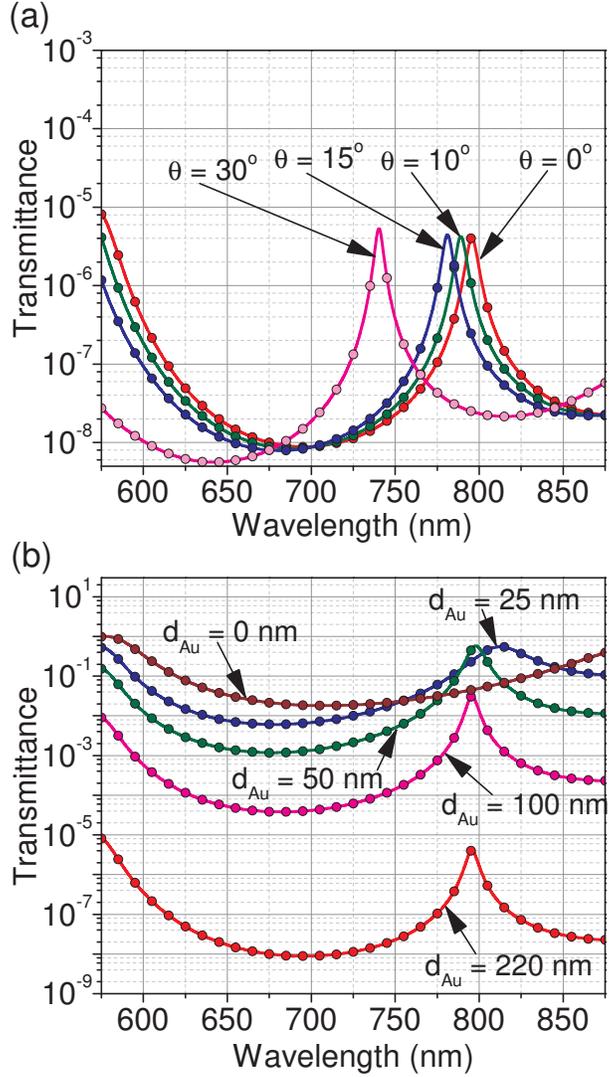}}
\caption{Dependences of the light transmission coefficients through
the optical structure ``metamaterial-metal film'': (a)~on the wavelength
and on the angle of incidence of light; (b)~on the wavelength and on
the depth of Au film~($d_{\text{Au}}$), normal incidence of light.
Solid curves correspond to the incidence of the light from the
side of~$\rm SiO_{2}$; dots show the reflection from the side of the
immersion oil.}
\label{fig_4}
\end{figure}

Conversely, in accordance with the reciprocity principle for planar
structures~\cite{ref_17}, the situation for the light transmission
coefficient~[see~Fig.~\ref{fig_4}] is always \textit{symmetric}
with respect to
the direction of incidence of the light; i.e., the transmission
coefficients for the light
incident~\textit{from opposite sides completely coincide with each other}.

Investigations of the behavior of the properties of the
structure~``metamaterial-metal~film'' upon variation of the film thickness
from~100~nm~to~220~nm showed that the shape of the reflection
curve~[see~Fig.~\ref{fig_2}(b)] changes insignificantly, whereas
the transmission
coefficient~[see~Fig.~\ref{fig_4}(b)] changes by more than
three orders of magnitude.

\section{\label{Section_04} Optical properties of an optical structure
``metamaterial-metal film'' with nanoapertures}

\begin{figure}[h!]
\center{\includegraphics[width = 84 mm]{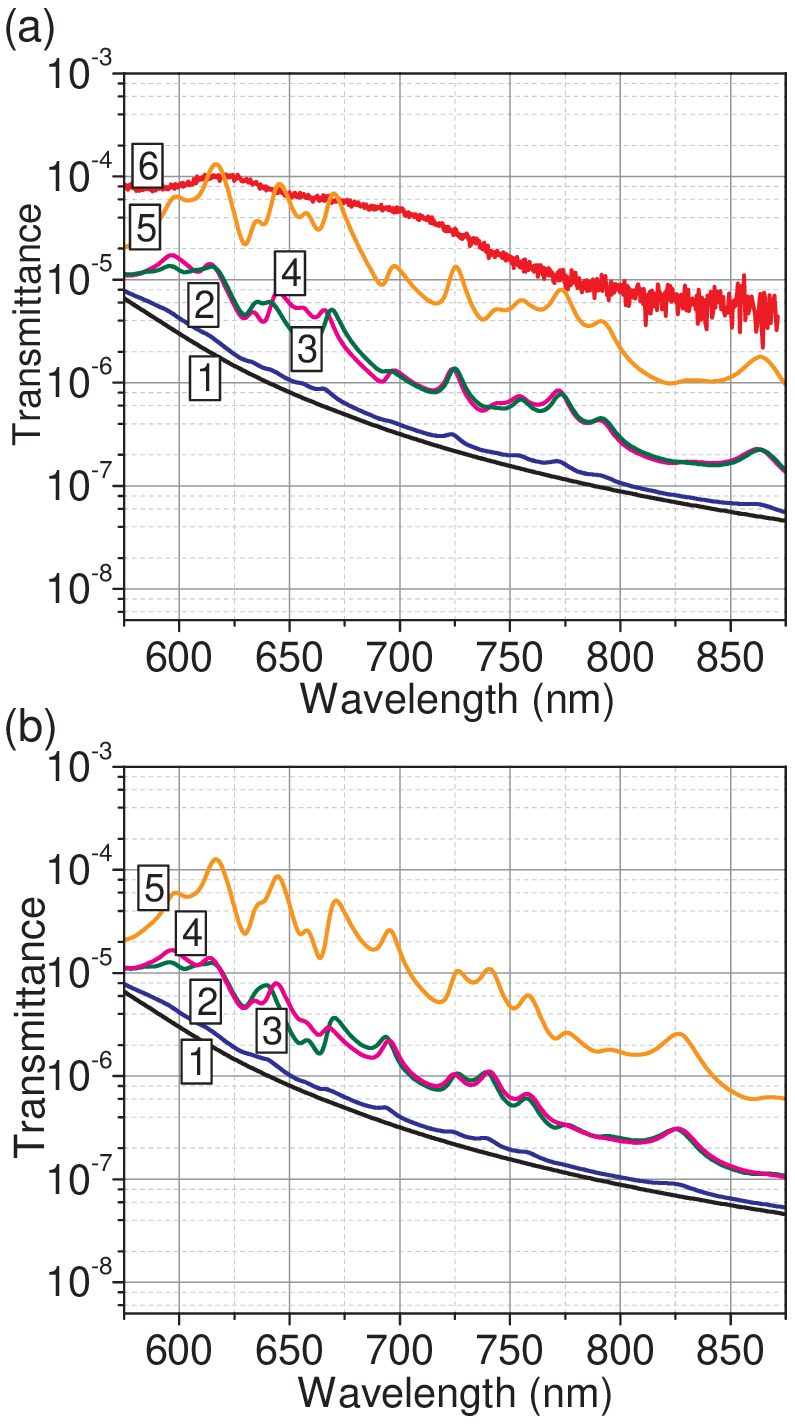}}
\caption{Dependences of the transmission coefficient through the gold film
with apertures but with no metamaterial on the wavelength of the incident
light for different geometries of nanoapertures: (a)~the light is incident
from the side of the~$\rm SiO_{2}$~layer; (b)~the light is incident from
the side of the immersion oil. Curves:~(1)~the case with
no~apertures~($d_{1} = d_{2} = 0 \text{ nm}$); (2)~apertures in the shape
of a cylinder~($d_{1} = d_{2} = 60 \text{ nm}$); (3)~apertures in the shape
of a truncated cone~($d_{1} = 60 \text{ nm}$, $d_{2} = 100 \text{ nm}$);
(4)~apertures in the shape of a truncated
cone~($d_{1} = 100 \text{ nm}$, $d_{2} = 60 \text{ nm}$);
(5)~apertures in the shape of a cylinder~($d_{1} = d_{2} = 100 \text{ nm}$);
(6)~experimental data from~\cite{ref_09}.}
\label{fig_5}
\end{figure}

Having clarified the behavior of the optical
structure~``metamaterial-metal film'' with no apertures, we consider now how
apertures in the gold layer affect the light transmission through
the structure. To do this, we used the finite element method, implemented in
the program~$\rm COMSOL^{\text{\textregistered}}$ (with a relative
calculation accuracy of no worse than~$10^{-3}$~at~resonance). Further, under
the transmission coefficient of light we understand the energy flux through
the area of one period of the optical lattice structure normalized to
the flux incident onto the same area. The light transmission coefficient
defined in this way is always smaller than unity. Figure~\ref{fig_5}~shows
the light
transmission coefficient through a gold film with different apertures but
with no metamaterial.

It can be seen from this figure that there is qualitative coincidence
between the experiment of~\cite{ref_09} and the numerical calculation for
cylindrical apertures with a diameter of~100~nm. Furthermore, there is
no resonance transmission related to the periodicity of the lattice, as that
observed in the experiment by Ebbesen~\cite{ref_01, ref_06}.

\begin{figure}[h!]
\center{\includegraphics[width = 84 mm]{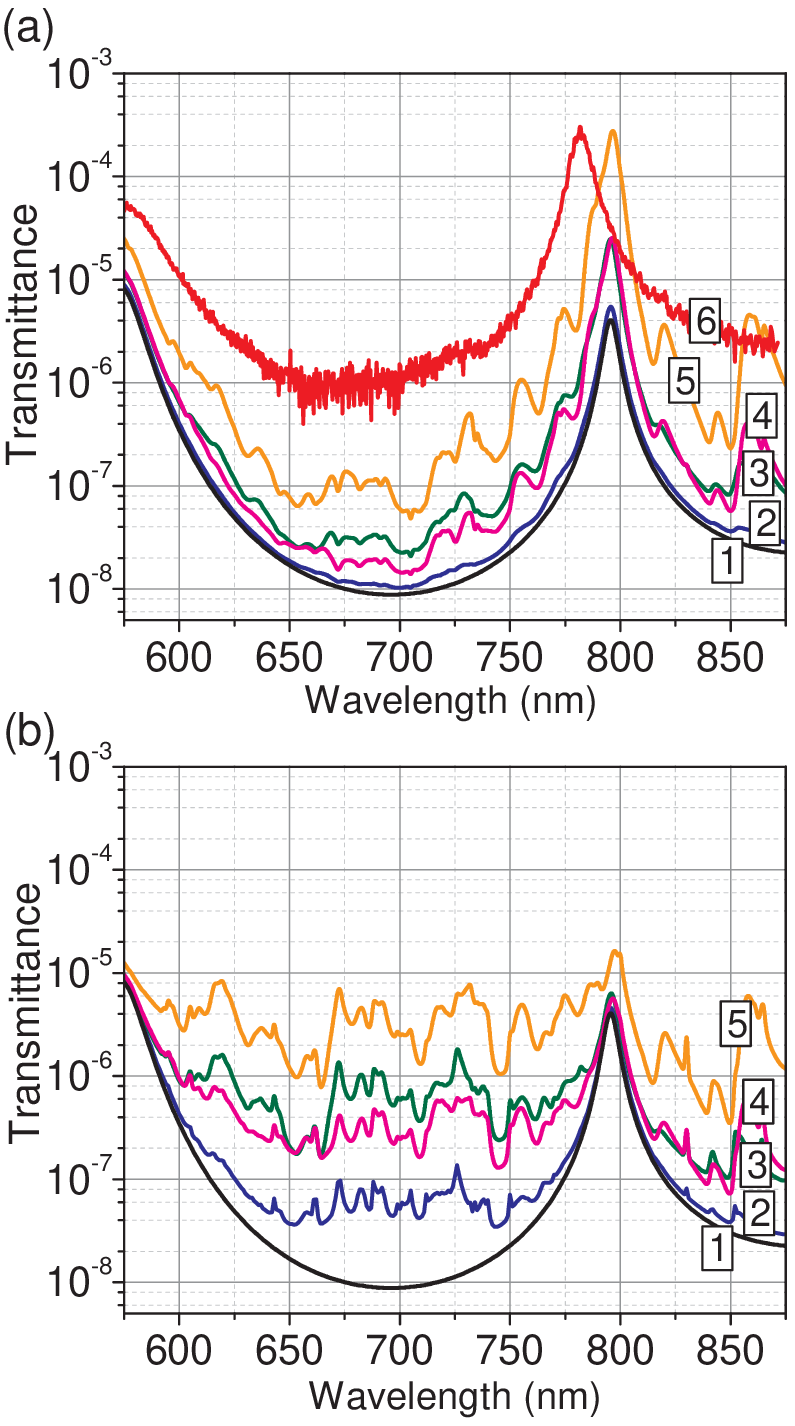}}
\caption{Dependences of the transmission coefficient through the optical
structure ``metamaterial-metal film'' with apertures on the wavelength of
the incident light for different geometries of apertures
((a)~the light is incident from the side of the metamaterial;
(b)~the light is incident from the side of the gold film):
(1)~the case with no~apertures~($d_{1} = d_{2} = 0 \text{ nm}$);
(2)~apertures in the shape of a cylinder~($d_{1} = d_{2} = 60 \text{ nm}$);
(3)~apertures in the shape of a truncated
cone~($d_{1} = 60 \text{ nm}$, $d_{2} = 100 \text{ nm}$);
(4)~apertures in the shape of a truncated
cone~($d_{1} = 100 \text{ nm}$, $d_{2} = 60 \text{ nm}$);
(5)~apertures in the shape of a cylinder~($d_{1} = d_{2} = 100 \text{ nm}$);
(6)~experimental data from~\cite{ref_09}.}
\label{fig_6}
\end{figure}

In the case of a gold film on the surface of a metamaterial (the optical
structure ``metamaterial-metal film'' with nanoapertures), the situation
radically changes. Figure~\ref{fig_6}~shows the coefficient of the light
transmission
through this structure~[see~Fig.~\ref{fig_1}(b)].

It can be seen from~Fig.~\ref{fig_6}~that, in this structure, there arises
an anomalously high light transmission at the resonance
frequency~($\lambda$ = 796 nm). It is also seen that the shape of the obtained
transmission curve coincides well with the measurement data
from~\cite{ref_09}. The shift of the resonance of the transmission
coefficient can evidently be a consequence of factors such as (i)~the use
of the oblique incidence in the experiment, (ii)~deviations in the values of
the optical constants that were used for the description of gold,
and (iii)~errors of preparation of experimental specimens. In addition,
possible reasons can also be the distinction of the shape of the real aperture
from the shape of the aperture used in our model, and the occurrence of a
shell from gallium atoms on the aperture walls, which appears because
apertures were formed with the gallium ion beam. The thickness of the gold
film strongly affects the transmission coefficient. It is interesting to note
that the transmission curves for the two types of apertures in the shape of
a truncated cone, which differ only in the orientation, are almost
coincide~[see~Figs.~\ref{fig_5}~and~\ref{fig_6}]. We can conclude from this
that the transmission
is mainly determined by the effective volume of the aperture rather than by
its shape.

Also, we note that the numerically simulated curves of the transmission
coefficient in Figs.~\ref{fig_5}~and~\ref{fig_6}~have multiple maxima and
minima, whereas the
experimentally determined curves do not show such
features~\cite{ref_09, ref_10}. This difference is caused by interference
effects in the lattice of nanoapertures, since we used periodic boundary
conditions at the cell edges~[see~Fig.~\ref{fig_1}(b)]. At the same time, the
experiments of~\cite{ref_09, ref_10} were performed on a single nanoaperture,
and no interference phenomena were observed in this case. Results of
investigation of these interference effects and their possible applications
will be considered elsewhere~\cite{ref_18}.

Comparison of Figs.~\ref{fig_6}(a)~and~\ref{fig_6}(b)~shows that
the transmission coefficient
considerably depends on the direction of incidence of light. If the light is
incident on the structure from the side of the metamaterial, the light
transmission considerably increases (by two orders of magnitude) compared
to the case of an aperture in a metal film alone, whereas, if the light is
incident from the side of the metal film, no increase in the light
transmission is actually observed. Therefore, the structure
``metamaterial-metal film'' can serve as a basis for the creation of an
\textit{optical diode}, a device that transmits light in one direction and is
opaque in the opposite one. A detailed investigation of this optical diode
will be described in \cite{ref_19}.

\begin{figure}[h!]
\center{\includegraphics[width = 84 mm]{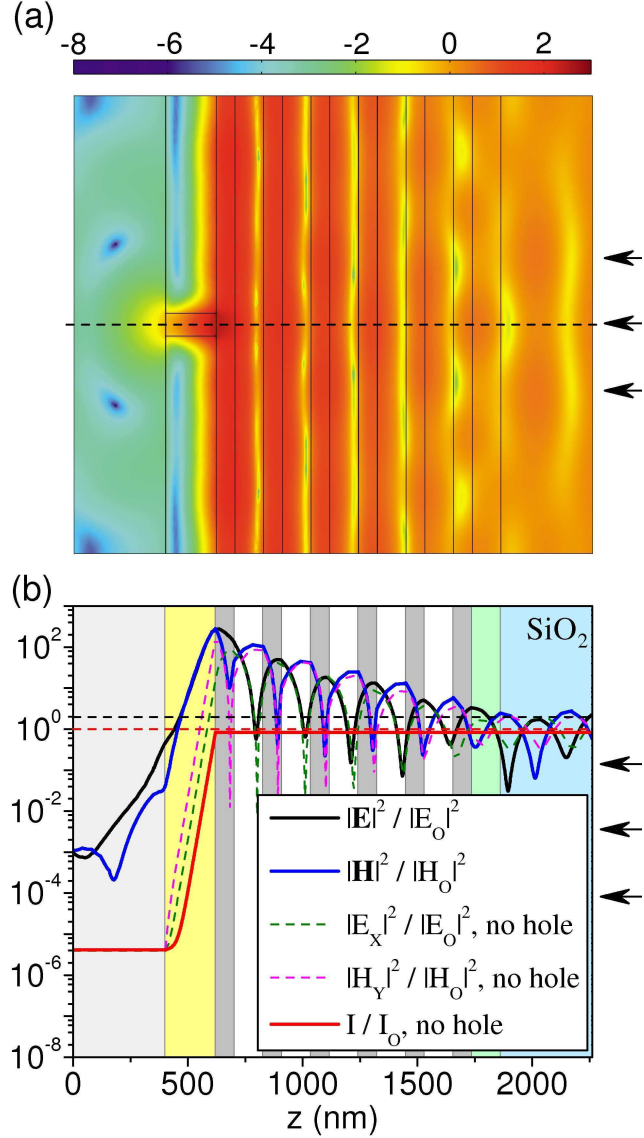}}
\caption{(a)~Distribution of the electromagnetic
field~($|\mathbf{E}|^{2}/|\mathbf{E}_{0}|^{2}$)~of a plane light wave
propagating through the optical structure~``metamaterial-metal film'' with
apertures in the case where the light wave is incident on the structure
from the side of the metamaterial at resonance~($\lambda = 796 \text{ nm}$).
(b)~The section of the two-dimensional distribution~(a)~along the dashed line;
the red solid line corresponds to the case with no aperture. A logarithmic
scale is presented.}
\label{fig_7}
\end{figure}

\begin{figure}[h!]
\center{\includegraphics[width = 84 mm]{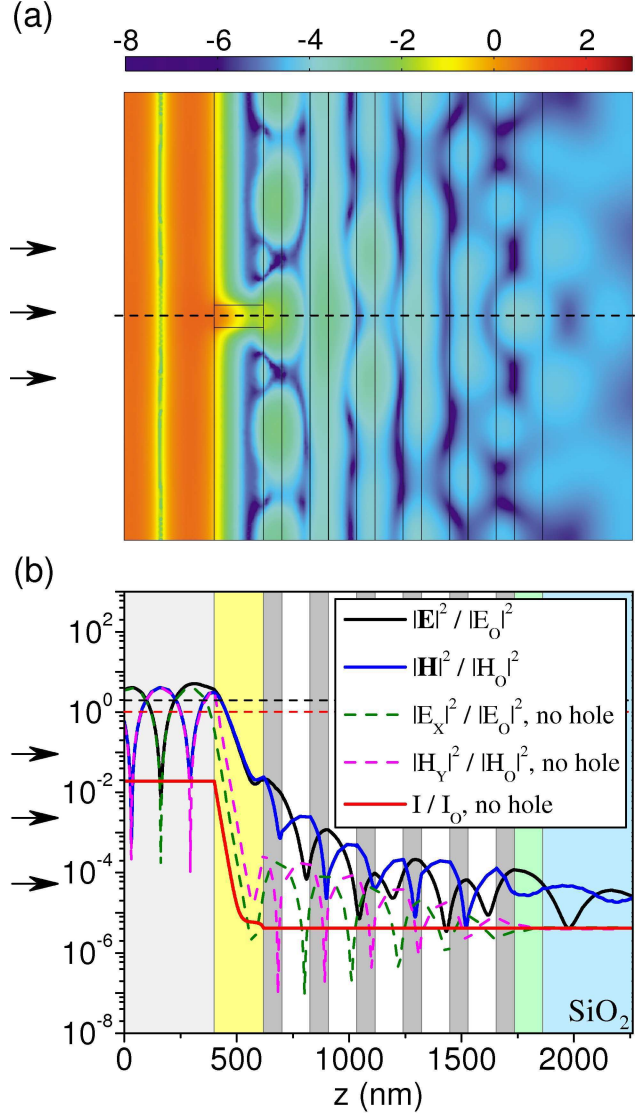}}
\caption{(a)~Distribution of the electromagnetic
field~($|\mathbf{E}|^{2}/|\mathbf{E}_{0}|^{2}$)~of a plane light wave
propagating through the optical structure ``metamaterial-metal film'' with
apertures in the case where the light wave is incident on the structure from
the side of the \textit{metal film} and at
resonance~($\lambda = 796 \text{ nm}$). (b)~The section of the two-dimensional
distribution~(a)~along the dashed line; the red solid line corresponds to the
case with no aperture. A logarithmic scale is presented.}
\label{fig_8}
\end{figure}

The picture of the light transmission through the aperture in the structure
under study becomes more clear from the consideration
of~Figs.~\ref{fig_7}~and~\ref{fig_8}, which
show the distributions of the
electromagnetic field~($|\mathbf{E}|^{2}/|\mathbf{E}_{0}|^{2}$)~in the cases
of the normal incidence from the top and from the bottom at
resonance~($\lambda = 796 \text{ nm}$).

As can be seen from~Fig.~\ref{fig_7}(a), when the light is incident from
the side of the
metamaterial, the amplitude of the field behind the aperture is comparable
with the amplitude of the incident wave.

Far away from the aperture, the field decreases as a spherical wave according
to a law $\thicksim 1/r$. In addition, the formation of a dipole directivity
pattern of the transmitted wave is clearly seen in~Fig.~\ref{fig_7}(a).
Figure~\ref{fig_7}(b)~shows the section of the two-dimensional distribution
along
the dashed line. The red solid line shows the intensity distribution along
the structure for the case with no aperture. The black and blue dashed lines
show the dependences of the squared electric and magnetic field components
along the structure for the case with no aperture. The corresponding
distributions for the case with the aperture are shown by the black and blue
solid lines. Figure~\ref{fig_7}(b)~shows how the magnitude of
the electromagnetic field
increases in the direction toward the gold film and an optical Tamm state is
formed on the surface of this film due to the constructive interference of the
light propagating in the metamaterial. It is also seen that the magnitude
of the field on the inner surface of gold at the center of the aperture
is higher than in the absence of the aperture.

If the light is incident from the side of the metal film, Fig.~\ref{fig_8}(b),
no enhancement of the field in front of the metal film occurs, and the
amplitude of the field behind the aperture is considerably smaller than
the amplitude of the incident wave. Therefore, in the
structure ``metamaterial-metal film'' with nanoapertures, the reciprocity
principle does not hold~\cite{ref_19}.

\section{\label{Section_05} Discussion of results}

In this work, using the finite element method, we have numerically simulated
the light transmission through a periodic lattice of nanoapertures in a gold
film deposited on the surface of a metamaterial. An effect of an anomalously
high light transmission has been revealed, which we associate with the
enhancement of the field at the interface ``metamaterial-metal film'' due to
the appearance of an optical Tamm state.

We also have shown that the optical structure ``metamaterial-metal film'' with
nanoapertures is an optical diode: the light transmission radically changes
(by two orders of magnitude) as the direction of irradiation of the structure
is reversed. Our numerical results agree well with the experimental results
of~\cite{ref_09} for a cylindrical nanoaperture with a diameter~of~100~nm.

Qualitatively, an increase in the light transmission through the nanoaperture
can be estimated in the case of interest by applying the Bethe-Bouwkamp
theory simultaneously with considering the enhancement of the local magnetic
field due to the occurrence of an optical Tamm state. Indeed, according to the
Bethe-Bouwkamp theory, the transmission coefficient of light at a wavelength
of~$\lambda = 796 \text{ nm}$~through a cylindrical nanoaperture with
a diameter of~$d = 100 \text{ nm}$ is
$T_{Bethe}=(64\pi^{2}/27)\cdot(d/\lambda)^{4}\thickapprox5.8\cdot10^{-3}$
(here, we use the normalization to the light flux incident on the aperture
cross section). The coefficient of transmission through the film with
apertures alone at a wavelength of~$\lambda = 796 \text{ nm}$~found by
numerical simulations is $1.7 \cdot 10^{-3}$
(see~Fig.~\ref{fig_5}; with the normalization to the flux incident
on the aperture
cross section). Qualitatively, these quantities coincide. The smaller value
obtained upon the simulation is seemingly related to a greater thickness of
the gold film~(220~nm).

In the case of the film with apertures on the metamaterial, the transmission
coefficient at a wavelength
of~$\lambda = 796 \text{ nm}$~is $1.4 \cdot 10^{-1}$~(see~Fig.~\ref{fig_6};
here, the normalization to the flux incident on the aperture cross section is
also used). That is, the use of the metamaterial leads to a~90-fold~increase
in the transmission coefficient. At the same time, the intensity of the
magnetic field in the metamaterial, which excites the magnetic dipole moment
of the nanoaperture (which, in turn, forms the transmitted field), is~50~times
higher than the intensity of the magnetic field on the surface of the
gold film. That is, there is a qualitative coincidence between these
quantities, which allows us to state that the extraordinary light transmission
in the case under consideration is related to an increase in the intensity
of the magnetic (as well as, electric) field. Thus, the transmission
coefficient of our structure can be estimated with the following simple
formula indeed:

\begin{equation}
\label{(2)} T = T_{Bethe}G,
\end{equation}

where $T_{Bethe}=(64\pi^{2}/27)\cdot(d/\lambda)^{4}$ is the Bethe-Bouwkamp
transmission coefficient, and~$G=|\mathbf{H}|^{2}/|\mathbf{H}_{0}|^{2}$~is
the enhancement of the magnetic field intensity at metal surface due to
appearance of Tamm state.

Therefore, in this work, it has been theoretically shown that the
enhancement of the field related to the occurrence of the optical Tamm state
of the electromagnetic field leads to a corresponding increase in the light
transmission through the nanoaperture.

\begin{acknowledgments}
V.V. Klimov and I.V. Treshin acknowledge financial support from
the Russian Foundation for Basic Research (grants \#\# 11-02-91065,
11-02-92002, 11-02-01272, and 12-02-90014) and from the Presidium of
the Russian Academy of Sciences. V.V. Klimov also thanks
the Russian Quantum Center for partial financial support.
The work of V.I. Balykin and P.N. Melentiev was supported by
the Russian Foundation for Basic Research, by the Program
``Extreme Light Fields'' of the Presidium of the Russian Academy of Sciences,
and by the Ministry of Education and Science of the Russian Federation.
Finally all authors thank Skolkovo foundation 
for substantial financial support.
\end{acknowledgments}

\end{document}